# On the principles of description of time and space relationships in frames of general relativity

S.I.Tertychniy *

Institute of Theoretical Physics,
University of Cologne,
Cologne, 50923, Germany

Time and space relationships...

## Abstract

The problem of the referring of space and time relationships between physical objects in a curved space-time is discussed. The basic notions of column and weak column that could constitute the basis for consistent general relativistic description of measurements of space and time relations are introduced and the corresponding equations are derived. A column of observers is proposed to realize still vague notion of 'non-inertial frame of reference' in case of curved space-time, the criterion distinguishing 'inertial columns' is suggested as well. A number of solutions of column equations is described. They are applied for calculation of gravitational, inertial (or gravitational-inertial in more general situation) and "external" forces and the strength of gravitational field in several cases of physical interest. In particular, the 'general relativistic' version of Newton's law of inverse squares is derived in case of the interaction of test particle with a static central symmetric body.

## 0. Motives and objectives

Situation that characterizes general relativity (GR) may seem somehow strange – at least in one respect. Indeed, one can not disregard the achievements of formal mathematical branches of theory that deal with objects that are typical of differential geometry and topology, theory of differential equations and other fields of mathematics and mathematical physics. There are also no doubts about applicability of the formal results of general relativity to the physical world – certainly, within certain limits, as the case is with any theory.

---

At the same time, one gets the feeling of dissatisfaction as soon as one starts to analyze fundamentals of practical methods of interpretation of the formal results of the theory connecting them with real measurements. Without any intention to question correctness of the main *numerical* results of experiments that nowadays form a sufficiently reliable criterion for distinction of GR as the most plausible among the other gravitation theories, one has to admit, anyhow, that the principal aspect of what one is tempted to name a "physical interpretation of general relativity" is hardly characterized by clarity, completeness and consistency.

We can illustrate it with the following example.

The effect of light deflection near a massive body (in practice a change of apparent position of a star or quasar when observed close to the solar limb, for example, during eclipse of the Sun) is well known, explained by the theory and can be proved by an experiment within the achieved accuracy (up to 0.1%, see [1]). Still, the problem concerning with a comprehension of this result remains. Indeed, according to GR, a light beam is represented in 4-dimensional space-time by the "straightest" world line, *i.e.* by a geodesic. So what is the "standard" more straight than geodesic from which light beam "deflects"? Indeed, it is quite obvious that when a star is observed near the solar limb its ray comes from a "wrong direction"; still if passing close to the Sun it *becomes* "curved", then something should *remain* "straight", provided our words are of any definite sense in this case. (The only alternative possibility here is to abandon the very notion of *a constant direction in space* but it is not a simple matter to make such a step.)

Certainly, one can stick to the obvious interpretation that says that the light deflection corresponds to displacement of the *visual* position of its source on the celestial sphere, while the "actual" direction to it coincides with its *initial* position (and could be defined in such a way). Then the "true" straight line from which the light beam deviates can be determined. Indeed, it is clear that no changes occurred to the observed star or to laboratory during the eclipse and only some intermediate space region through which light passes becomes distorted. In the outer regions the 'straight lines' determining the direction to the source must remain the same.

Anyhow, such an approach is too limited to be universal one since it assumes in fact the existence of some additional "always fixed" light sources that enables to refer the position of the distorted one. This was always being satisfied in practice but only until the gravitational field is sufficiently weak and accuracy of experiment is sufficiently low.

Nowadays, however, it is likely that the latter condition is close to be violated but nevertheless it seems to be not the most important point. The reason is, of course, clear.

For, let us imagine – by a way of *gedanken* experiment – that somebody find himself on a stable orbit in a close – although safe – proximity to a large number of black holes; let the motion of these black holes be stable, almost periodic and, finally, let their number be so large and their apparent motion be so sophisticated that all the time a prevailing part of celestial sphere of observer is covered by a projection of areas with extremely strong gravitational field and is drastically distorted in a practically unpredictable way. In such a situation, the locations of images of remote stars certainly can not be firmly settled on the celestial sphere and applied for definition of a standard of constant directions in space.

One has to accept without further arguments that in a general case, when the space-time is not stationary and its curvature is essential, observation of "fixed" remote stars cannot be regarded as a satisfactory method of observer's orientation that should enable him to recognize his own revolution. Similar statement is also valid in the less striking but much more close to reality case when the desirable accuracy of orientation exceeds the "mean" value of light deflection due to the nearest to observer massive objects.

All the more, the references to remote stars cannot be used during consideration of principal questions of observer orientation.

It is clear that a light beam may not preserve its initial direction along sufficiently long parts of rays in a general case due to gravitational deflection although it transports tangent vector paralelly along own worldline. All the more, we must then question a physical reasonability of various "local frames of reference" based on the parallel (or, its direct generalization, Fermi–Walker) transport of orthonormal frame along worldline as a method of organization of oriented non-rotating frame of reference since they represent not more than a 'deteriorated' version of the transport of tangent vector along a null geodesic.

Independently of the above qualitative speculation, the following paradox additionally argues in favor the same conclusion. It is often claimed that direction control can be maintained with the help of gyroscopes described just by the Fermi-Walker transport [2]. At the same time it is well known that a free gyroscope is subjected to a precession resulting from its interaction with gravitation field ([3],[4], p. 186) or even merely from its motion as far as it is relativistic (Thomas's precession [4], p.112) although we have yet no experimental evidences of these rather weak effects. The first and the second treatments of the ability of a gyroscope to keep the constant direction clearly contradict one to another, but a crucial point is that before any discussion of a precession (which can be unambiguously interpreted at least in case of flat space-time and, thus, can be considered as a real subject of experimental investigation) it is necessary, in case of curved space-time, to specify with respect to *what* direction this precession occurs, in what sense this direction is constant and how it is being maintained (in Minkowski space such a problem does not arises, of course). There are all reasons to think therefore that a vicious circle is inevitable here.

Apparently the problem of observer's orientation in space needs at least a more accurate treatment.

Alongside difficulties related to fixation of a method of 'directional' control over observer's orientation in a strong gravitational fields there are similar difficulties with the 'positional' orientation.

Idealized measurement of *distance* with the help of "rulers" or "non-elastic cords" is rejected already by special relativity (SR) which puts forward the principle of *light ranging* instead. Certainly, it can be also applied in case of curved space-time, still in this case it will imply answers to a number of additional questions.

For, let us imagine that four observers sending each other light signals are moving in such a way that at any moment the distance determined as half time of travel of light signal multiplied by velocity of light remains constant between any two of the four [1]. We have what could be called a "stiff tetrahedron". The question is: Is it possible for *the fifth*

---

[1] It is far from obvious that the distance measured in such a way will be symmetric, that is if the first observer supposes that the second one is always at one and the same distance from him,

observer to move in such a way that he would remain at the fixed distance from any of the four? It is quite obvious that in general case the answer will be negative.

All the more, it is impossible to provide a continuous "media of observers" in such a manner that each two points of this media were stationary in respect of each other, *i.e.* the distance between them measured by light ranging remains constant. Then how can one access, for example, the value or the variation of value of velocity of a test particle moving through the set of observers if all of them move one relative to others? Indeed, we have *no* other basis to refer the motion of observers that should serve this purpose *themselves*.

Another question may sound like this. An observer will never receive a light signal reflected from the black hole. Whether does it mean that the black hole is situated at an infinitely long distance from any observer outside the horizon? It would be rather strange because, in particular, a test particle at a free fall reaches the horizon of a black hole in a finite time.

One shall agree that a measurement of distance by light ranging requires additional analysis and possible refinement.

Thus, we are dealing with a rather vague principal approach to such basic notions as space orientation and application of quantitative assessments for description of reciprocal location of interacting objects. This draw-back is sufficiently serious for a theory that pretends to provide a consistent description of space and time relationships between physical objects and events to try to overcome it.

Now let us come down however to the earth and be slightly closer to reality. It is easy to see, of course, why, contrary to our speculations, in practice the situation seems to be in no way dramatic, the experiments are well interpreted and yield reasonable results that agree with what one expects. The thing is that in most cases well founded and reliable physical interpretation from the *special* relativity is used (and, as a rule, it is quite justified [2]). The effects typical of a curved space are considered only as minor perturbations of the basic pattern of a flat space, meanwhile *the physical meanings* of the basic notions is borrowed (often, in fact, together with the Minkowski space itself) directly from SR without any corrections, no matter that it is not sometimes explicitly declared. For example, the PPN method is nothing else but the application of the flat geometry to a slightly curved space-time. This is usually sufficient for description of effects occurring within the Solar system because of their smallness; meanwhile a questionable embarrassment of a hypothetical observer surrounded by black holes is not regarded as a sufficient reason for revision of such a convenient and habitual methodology. Even if we study a process in a strong gravitational field we usually need the physical interpretation (*i.e.*, essentially, the description of relevant measurements) only on the stage of consideration of its far consequences (for example, spreading radiation) and their influence to earth-based instruments, where such effects are

---

the second one will be convinced in the same thing. Here we ignore this additional problem.

[2]) The exceptions are however possible: it is proved in the work [5] that more careful than usually analysis of the connection between the geometry and the results of measurements in the case the light deflection effect yields the prediction that differs from the classical one already in the first approximation with respect to $m/r$. The corresponding discrepancy is expected to be on a level of recent measurement uncertainty.

extremely relaxed and the perturbation approach based on SR-like interpretation is quite sufficient.

Still, there is of course situations that cannot be understood within the frames and notions typical of a flat space-time. The most evident example is cosmological phenomena. The complex problems of cosmology [6] are not a subject of this paper but one remark is worthy to be made. We speak about comprehension of cosmological expansion of the Universe as a recession of galaxies. The very term of 'recession' reflects an intuitive attempt to apply the flat space-time geometry for interpretation of the essentially non-stationary process. In this case however we observe the 'recession' of objects that are situated from an observer at a distance comparable to space-time "radius of curvature", and the approach based on perturbations and, most important, keeping the basic interpretations borrowed from SR cease to be sufficient. In frames of notions of Minkowski space geometry it is impossible, for example, to fit the spatial homogeneity and central symmetry of the Friedmann model with the obvious fact that the 'recession' of galaxies, if any, generates distinguishable directions *outwards*. (It becomes more clear if one considers three 'cosmologically immobile' objects situated along common "straight line". Indeed, the observer living on the 'middle' object will be sure that both of the 'remote' objects are moving away from him in opposite directions. But when these 'remote' objects are chosen so far (form the middle one) that they turn out to be close between themselves (case of closed model) and, furthermore, tend to merge, the observer comes to incomprehensible situation since he must to admit that the single object moves at the same time in two opposite directions, both away of him.)

Despite the fact that there is perhaps only a limited number of such type examples, it is logical to assume that principal basis of 'physical interpretation' may be equally important for both GR and, for example, quantum theory, where principles of object-apparatus interaction enter the very core. The difference with our case is perhaps of more quantitative than principal character. Regretfully, a lack of satisfactory physical interpretation results sometimes in unjustified domination of formal approaches perhaps mathematically perfect but missing in certain cases a physical content of things; as a result their application can hardly be successful. As an example we can refer to the problem of gravitation field energy. Here the investigation of *local interactions* of objects and fields accompanied by the *mutual energy exchanges* (which clearly exist and are of the most physical interest) is often replaced by the searching for some formal conserved semiglobal integrals or even more abstract exercises whose relation to a physical energy notion is rather questionable.

What are, however, the ways to achieve a consistent "physical interpretation" of general relativity? One might suggest at least two of them. The first way is merely to "guess" how it should be. It will not be followed here however. Secondly, one may try to analyze how to realize the principle of correspondence between GR and SR and what aspects of SR physical interpretation might be used unchanged in GR. Having distinguished these elements, one could further apply a logic and common sense to extend it to a more or less comprehensive picture.

We intend to undertake a (certainly restricted) attempt of the latter kind here and suggest some methods that could be useful in the solving of problem of measurements of space and time relations and investigation of fields dynamic.

## 1. Concept of column

At the very beginning, one must decide what principles of the special relativity describing space and time relationships can be also applied to a curved space-time. The first principle is quite obvious and specifies the geometrical interpretation of the notion of time:

**1**. *Time measured by any observer regardless of nature of his motion and the state of space-time coincides with the value of interval along his world line between corresponding events.*

It seems likely to exist no experimentally or theoretically maintained reasons that would force us to be uncertain with respect to such property of a local time.

It can be noticed that the independence of the 'course of time' on acceleration has been confirmed in experiments with the rate of decay of muons [7], the acceleration achieving $10^{18}$**g**. There exist estimates [8] predicting a measurable distinction (up to 10%) of the muon's decay-rate from expected value in extremely strong magnetic field in vicinities of neutron stars although there is no observational evidence of existence of such a discrepancy. It should be emphasized, however, that the change of behavior of a clock of certain type does not mean generally speaking the change in behavior of the time itself. Such a conclusion might be put forward only if all the clocks change their course due to some general cause which usually corresponds to some universal principle (such, for example, as the Lorentz invariance of physical laws implied by SR).

Next we have to describe basic properties of our principal probe, a light beam.

**2**. *Regardless of nature of observer's motion and the state of space-time in an infinitely small neighborhood of observer light propagates in a constant direction with invariable velocity that is numerically equal to the known world constant.*

This postulate may be interpreted as definition of distance between two *infinitely close* observers: it equals half time of light signal travel multiplied by world constant of light speed.

We have no reasons to doubt this statement as well; furthermore, the principle of correspondence with SR dictates such an assumption unambiguously. It is important since otherwise we would be unable to provide any consistent background for the notion of distance in space.

One may face also the problem of defining of notion of infinitely close observers; still we believe that this question is rather clear on a level of non-quantum physics and controversy is not likely to be expected in this point; thus, we are not going to discuss it. Only point to note is that the procedure of determination of distance between two *infinitely close* observers with the help of light ranging will be briefly called "nearby ranging" or simply "ranging"; the same procedure for observers situated at a *finite* distance one from another will be called occasionally "finite ranging".

It seems also reasonable to make a remark concerning the sense of "constancy of direction" of light beam "in infinitely small neighborhood" of observer. This means that any two infinitely close observers detecting the same light beam may assume that it has *the same space direction* with a precision up to *the first* order of their mutual distance. In other words, the angle of light deflection (no matter that it has not been defined by us by this point) is in the worst case of the *second* order with respect to distance. This property distinguishes the zero-rest-mass particles (photons) from non-zero-rest-mass ones. In the later case, a deflection angle may by of the first order with respect to distance as the evident example of a free orbital motion in Schwarzschild spacetime shows.

Finally, any observer feels himself to be immersed into 3-dimensional space. We have the following statement.

**3**. *Space in a neighborhood of any observer is locally Euclidean. This implies the well-known connection between distances (measured by means of nearby ranging) and angles which is expressed by cosine theorem and yields in fact the definition of measurable angle.*

Local Euclidean property is a usual requirement, it testifies the absence of conical singularities in space.

The stated principles, although not restrictive enough, are sufficient for the setting-up the mechanism of space and time relationships measurement. For this purpose we introduce the central concept of *the column of observers.*

**Definition 1:** *Column of observers (or simply column) is a set of observers that continuously fill a certain open domain of space and move in such a manner that any two infinitely close observers remain at the same distance one from another during their entire history.*

We may say that the column realizes a *locally* absolutely stiff "media" of observers. It has been mentioned above that it is impossible for every pair of more than 4 observers to be mutually immobile in a general case. Here we use however a *local* immobility (immobility of infinitely close observers) only and such a requirement is considerably weaker (although some problem still remains, as we shall see below).

As for the term 'column' itself [3]) (the English version of original Russian 'строй'), we mean to draw analogy with the military meaning of this word: a group of organized individuals engaged in co-ordinated evolutions keeping all the time the same fixed distance one from another.

A column is meant in particular to replace SR inertial frame of reference in case of curved space-time. Still, in case of column no restrictions are *a priori* applied to eliminate rotation of this locally stiff "media" of observers as whole as well as its acceleration (although some further secondary boundary or asymptotic conditions can serve such purposes). Thus a column better corresponds to a notion of non-inertial frame of reference (that, by the way, has not been satisfactorily defined yet).

Notice also that the property of light beam to keep locally its space direction does not enter the column definition.

Global synchronization procedure is not determined by us. Only a local kind synchronization is possible and comprehensible. It is interpreted in the following way. If the first observer who is at an infinitely short distance $\delta l$ from the second one at the moment $\tau$ of his own time received a light signal from the second observer, he may assume that the signal was sent *by the second* observer at $\tau - \delta l/\mathbf{c}$ of the *first observer's* watch. In a general case synchronization does not mean anything more. Nevertheless, as we shall see below, this is quite sufficient to determine *the velocity* of particle moving through the space served by column, its *momentum* and, finally, *the force* that acts on it.

---

[3]) It seems impossible to use any derivative of the expression *frame of reference* which (in case of curved space-time) is irreversibly compromised by innumerable inappropriate applications.

We shall consider that any two observers of the column do *not* move in relation to the other. This is the definition of column for infinitely close observers, for others this will be the definition of reciprocal immobility itself. We shall note, however, that "finite" ranging may yield *varying* values of time of signal travel for such defined reciprocal immobile observers even in case of flat space-time.

It would be noticed here additionally that for each two finitely separated observers one can make a file of (infinitely) large number of reciprocally (infinitely) close observers of the column and sum up (integrate) respective constant distances; thus one obtains the length of the curve (or "cord") connecting observers. Next one can find minimal length along all the possible curves and regard it as a distance between them.

Thus, one may try to integrate the simple and, in our opinion, natural principles **1** - **3** that are borrowed from special relativity and specified to the case of curved space-time into a procedure that could yield a general concept of physical time and distance, *provided* the column can be constructed for a given space-time.

Unfortunately, there are simple examples showing that in the case of more than 2-dimensional curved space-time the columns may not exist. Then somewhat less restrictive construction than a column as it has been defined above can be put forward. One of natural possibilities is realized by the following

**Definition 2:** *Let the (open domain of) space-time be foliated by a family of time-like hyperplanes and let each of them, in turn, be foliated forming the family of 2-dimensional timelike planes. Further, let a column be constructed over every such 2-plane determining moving rigid line. Families of rigid lines from a chosen hyperplane form rigid planes, provided the distance between every two neighboring rigid lines (not individual observers!) remains constant. Similarly, a family of foliating hyperplanes forms a weak column if the distance between any two neighbor hyperplanes is a constant.*

In other words, we stiffly fix the longitudinal position of observers along some lines and transversal distance between these lines and planes constituted of them but allow line to slip with respect to neighboring ones if it is inevitable due to features of space-time geometry. Of course, the (specific) velocity of such a relative motion of neighboring lines per unit distance between them may (and must) be determined and it is an important characteristic of a weak column.

All the same, there are some reasons to argue that construction followed from definition **2** is perhaps not universal and maybe there exist more advantageous versions of the weakened column notion. In any case, however, they have to follow the principle *to design a construction as stiff as possible keeping at the same time a total control over remaining intrinsic motions of its constituents, if any*. It is a continuous "media" of observers (families of time-like world lines) that *has to be used as the primary basis for the referring of space and time relations* and various kind vector frames must be considered as secondary objects only. On the other hand, the class of *all* timelike congruences is too wide and can not serve the generalization of the notion of inertial (and even non-inertial) frame of reference typical for SR since it does not admit corresponding flat-space limit.

It seems worthwhile to emphasize once more that in general case it might be impossible for a set of observers to be mutually immobile even in a local sense but it is certainly possible to *reduce their mutual motions to minimum*. If it has been achieved and if the remaining mutual motions of observers are properly described we shall have the sufficient desirable facility for referring of the motions of other physical objects and determination of variations of fields.

Despite of the latter remarks, one has to mention, however, that the concept of a "strong" column (Definition **1**) has a rather wide field of applications. We shall focus mainly on it in these notes.

## 2. The column equation

Let us assume that space-time (or its open domain) is covered by a single chart with coordinates $x^\alpha, \alpha = 1, 2, 3, 4$. Suppose further that we have a congruence of curves

$$x^\alpha = X^\alpha(t, u^a)$$

where $t$ is non-degenerate parameter along them; parameters $u^a$, $a = 1, 2, 3$ identify individual curves of the congruence. Let us suppose also that they are time-like:

$$\|\dot X\|^2 \equiv g_{\alpha\beta}\dot X^\alpha \dot X^\beta > 0 \tag{1}$$

and $t$ increases to future direction along them. A dot hereinafter denotes a derivative with respect to a parameter along curves, in our case $t$. Derivatives with respect to parameters $u^a$ will be denoted by Latin subscripts through comma. We assume $g_{\alpha\beta}$ signature to be equal to $(+ - - -)$.

Column differs from all other possible timelike congruences by the property to keep a constant distance measured by the light ranging between any two infinitely close observers along their worldlines. We shall derive now the equations realizing this property.

Let us choose any curve of congruence, and call it for a time a "basic observer". We shall disclose the definition of a column over the basic observer and any other infinitely close "neighboring" observer. One has the following equations:

$$\text{basic curve equation} \qquad x^\alpha = X^\alpha(t, u^a),$$
$$\text{neighboring curve equation} \quad x^\alpha = X^\alpha(t, u^a) + X^\alpha_{,b}(t, u^a)\delta u^b.$$

Here $\delta u^b$ is an infinitely small increment of parameters that differs the neighboring curve from the basic one, the second equation being valid up to terms linear in $\delta u^b$.

Let us assume that light signal reaches neighboring observer (and is reflected by him) at a certain value of parameter $t$; we shall not introduce a special notation for it. In proximity of events of light signal emission/reception we have the following equation of the basic observer's world line (accurate up to the first order of $\delta t$)

$$x^\alpha = X^\alpha(t, u^a) + \dot X^\alpha(t, u^a)\delta t.$$

Here $\delta t$ is a small (order of $\delta u^b$) variation of parameter along the path. One obtains the representation for components of infinitely small 4-vector $\delta x^\alpha$ connecting a "running" point at the basic curve and a signal reflection point at the neighboring curve:

$$\delta x^\alpha = \dot{X}^\alpha(t, u^a)\delta t - X^\alpha_{,b}(t, u^a)\delta u^b.$$

Suppose the light probe signal is emitted or reflected signal is received just at the moment $t + \delta t$. Then this vector shall be null, *i.e.*

$$0 = g_{\alpha\beta}\delta x^\alpha \delta x^\beta = g_{\alpha\beta}\dot{X}^\alpha \dot{X}^\beta \delta t^2 - 2g_{\alpha\beta}\dot{X}^\alpha X^\beta_{,a}\delta u^a \delta t + g_{\alpha\beta}X^\alpha_{,a}X^\beta_{,b}\delta u^a \delta u^b. \qquad (2)$$

Due to (1), this quadratic equation with respect to $\delta t$ is not degenerated. It is clear from its geometric origin that two corresponding roots must be real; we shall denote them $\delta t_+$ and $\delta t_-$. The proper time $\delta \tau$ measured along the basic curve between events of emission and reception of light signal is, most shortly, determined by the equations

$$\delta \tau^2 = g_{\alpha\beta}\dot{X}^\alpha \dot{X}^\beta (\delta t_+ - \delta t_-)^2 = \|\dot{X}\|^2 \left[(\delta t_+ + \delta t_-)^2 - 4\delta t_+ \delta t_-\right].$$

From (2) one can infer

$$\|\dot{X}\|(\delta t_+ + \delta t_-) = 2g_{\alpha\beta}\dot{X}^\alpha X^\beta_{,a}\delta u^a/\|\dot{X}\|,$$
$$\|\dot{X}\|^2(\delta t_+ \cdot \delta t_-) = g_{\alpha\beta}X^\alpha_{,a}X^\beta_{,b}\delta u^a \delta u^b$$

and the square of the proper time interval turns out to be equal to

$$\delta \tau^2 = 4g_{\alpha\beta}\dot{X}^\alpha X^\beta_{,a}\delta u^a \cdot g_{\mu\nu}\dot{X}^\mu X^\nu_{,b}\delta u^b/\|\dot{X}\|^2 - 4g_{\alpha\beta}X^\alpha_{,a}X^\beta_{,b}\delta u^a \delta u^b$$
$$= 4(g_{\alpha\beta}g_{\mu\nu} - g_{\alpha\mu}g_{\beta\nu})\dot{X}^\alpha \dot{X}^\mu X^\beta_{,a} X^\nu_{,b}\delta u^a \delta u^b/\|\dot{X}\|^2$$
$$= -8g_{\alpha\mu}g_{\beta\nu}\dot{X}^{[\alpha}X^{\beta]}_{,a}\dot{X}^{[\mu}X^{\nu]}_{,b}\delta u^a \delta u^b/\|\dot{X}\|^2.$$

In accordance with the sense of determination of distance by means of nearby light ranging in the system of units where light speed **c** equals unit the distance between neighboring observers is to equal just $\frac{1}{2}\delta\tau$. It shall be constant along the entire basic curve. Thus the functions

$$h_{ab} = -2g_{\alpha\mu}g_{\beta\nu}\frac{\dot{X}^{[\alpha}X^{\beta]}_{,a}\dot{X}^{[\mu}X^{\nu]}_{,b}}{g_{\rho\sigma}\dot{X}^\rho \dot{X}^\sigma} \qquad (3)$$

shall not depend on $t$ and

$$\partial_t h_{ab} = 0. \qquad (4)$$

With respect to transformations of parameters $u^a$ the components $h_{ab}$ constitute a symmetric tensor. It is evidently *the metric tensor* for a local distance measured by nearby ranging. Distance element is determined by the quadratic form $h_{ab}\delta u^a \delta u^b$. As opposed to metric $g_{\alpha\beta}$, tensor $h_{ab}$ will be called "physical metric" below.

Physical metric is to be positively determined. Otherwise, for a certain choice of $\delta u^a$ in accordance with the above formulae we would obtain $(\delta t_+ - \delta t_-)^2 \leq 0$. But this would mean, with inequality, that equation (2) has complex roots and has no real ones, *i.e.* the null cone with the vortex on the neighboring curve has no intersections with the basic curve that is infinitely close to it, or there is an intersection only with one cone sheet with degeneration of physical metric, in case of equality. Such a situation is impossible for timelike smoothly parameterized congruence. Thus, if condition for timelikeness (1) is observed, physical metric will yield positively determined quadratic form.

Obviously, the declaration of an observer a basic one has been used only to make the above derivation more demonstrative. In final equations no curve is distinguished in any way. Any curve from the congruence that satisfies equations (3), (4) can be used as a basic observer world line.

We shall call equations (3), (4) *the column equations*.

Notice that interval $\tau$ may always be chosen as a parameter along column curves. In such a case

$$g_{\rho\sigma}\dot{X}^\rho \dot{X}^\sigma = 1 \tag{5}$$

and it proceeds from (3), (4)

$$\partial_\tau \{g_{\alpha\mu} g_{\beta\nu} \dot{X}^{[\alpha} X_{,a}^{\beta]} \dot{X}^{[\mu} X_{,b}^{\nu]}\} = 0 \tag{6}$$

or

$$\partial_\tau \{g_{\alpha\beta} X_{,a}^\alpha X_{,b}^\beta - g_{\alpha\beta}\dot{X}^\alpha X_{,a}^\beta \cdot g_{\mu\nu}\dot{X}^\mu X_{,b}^\nu\} = 0. \tag{7}$$

Partial derivative with respect to $\tau$ can be replaced for the operator $\dot{X}^\alpha \nabla_\alpha$ and thus the column equations can be recast to completely covariant form. It is reasonable, however, to substitute the equations of the second order (6) by the system of their "first integrals"

$$g_{\alpha\mu} g_{\beta\nu} \dot{X}^{[\alpha} X_{,a}^{\beta]} \dot{X}^{[\mu} X_{,b}^{\nu]}(t, u^c) = -\tfrac{1}{2} h_{ab}(u^c). \tag{8}$$

In such a case, $h_{ab}$ can be *a priori* an arbitrary positive definite matrix, it is not specified beforehand. Still, it turns out that column equations are consistent only in case of quite definite selection of physical metric in fact, the scope of possibilities being determined first of all by the form of metric tensor.

System of column equations is obviously overdetermined: one has 4 unknown functions $X^\alpha(t, u^a)$ and 7 restrictions on them - 6 equations (8) and equation (5) (functions $h_{ab}$ depend on 3 or less variables and shall not taken into account in this count). It is true that equation (5) provides only natural parameter $\tau$ and can be omitted here. In the latter case, column equation acquires form (3) with an additional restriction of a positive definiteness of physical metric; in any case we still have 6 equations for 4 unknown functions.

Generally speaking, it is a good sign that the column equations are overdetermined. Otherwise, they should admit (at least locally) a solution depending on an arbitrary function of *three* variables that would be too much from the physical point of view. On the other hand overdetermined system may have no solutions at all.

The problem of compatibility of column equations and parameterization of their general solution seems to be one of the key current issues for column theory capability. Column equations are highly nonlinear (the fourth power with respect to various variables derivatives and the second power with respect to the same variable derivatives) and it is difficult to prescribe their general solution even in the most simple cases. However the *desirable* degree of a "power" of the set of their solutions can be characterized by the following criterion.

**Capability criterion.** *The general solution of column equations exists and depends upon **7** arbitrary functions of **one** variable.*

In such a case a column can be "leaned", at least locally, on any arbitrarily moving observer (four functions match his world line and three his directional orientation).

The latter alternative interpretation of criterion enables one to adapt it to the case of less then 4-dimensional space-time also.

If the capability criterion holds the set of all columns could be interpreted as a set of non-inertial frames of reference. Furthermore, one can conjecture that the column is in fact precise realization of somewhat vague notion of non-inertial frame of reference, at least in case of flat space-time where the capability criterion is likely to hold. We shall see in the next section how one could single out a subset of *inertial* frames from it.

Unfortunately we have no general results concerning the status of the capability criterion nowadays. There are some tractable examples in reduced dimensions, however, that *mutatis mutandis* obey it; they are partially outlined below.

It is worthwhile to notice that observers constituting column – according to the definition of local reciprocally immobile observers which has no relation to geodetic property – does not generally speaking follow geodesics (although in special cases it may occur, especially if there are symmetries of space-time). This means that free (geodesic) motion is not generally an "inertial" motion and hence some *gravitational force* interferes in it. Of course, in that cases column's observers are forced to use some engines to resist the gravitational force and maintain their specific positions.

It is easy to see that in Minkowski space-time every congruence of parallel timelike geodesics (constituting jointly the inertial frame of reference) is at the same time a column with a flat physical metric. However, not every column with flat physical metric is realized in such a way as an example of uniformly accelerated column (serving just the 'Rindler universe' $|t| < x > 0$) shows. There are also columns that correspond to non-zero curvature of physical metric even in the flat space-time and the uniformly rotating column yields the simplest example of such a sort. But it should be better to consider particular examples separately.

### 3. Examples

Here we present a number of simple but nontrivial examples of "strong" columns. Some their applications are developed, several of them being of a certain interest in their own rights. Almost every subsection below is worthy of a separate paper and we can give only principal formulae and minimal discussions here.

**3**.1 RADIAL MOTION IN CENTRAL SYMMETRIC STATIC SPACE-TIME

Let us consider 2-dimensional space-time $\mathcal{M}$ endowed with metric

$$ds^2 = F(r)dt^2 - G(r)dr^2 \tag{9}$$

It can be considered in particular as a trace of metric of a general 4-dimensional static spherical symmetric space-time onto its radius, motions along it and in $\mathcal{M}$ being governed by the same equations. In case of the latter interpretation, we also assume the coordinate $r$ to be defined through the area of sphere $r = $ constant by requirement that it should be equal to $4\pi r^2$. This assumption specifies $r$ unambiguously. At the same time $t$ coordinate is determined up to a shift $t \Rightarrow t+$constant.

In order to obtain the relevant form of column equations one may concretize the general equations of Sec.**2** to the case of metric (9) but it turns out to be more convenient to reproduce their derivation. Let the word lines of observers constituting column be described by equation

$$t = T(r, l) \tag{10}$$

where $l$ is the parameter enumerating observers such that the distance between a pair of them, labeled by $l$ and $l + \delta l$ parameter values respectively, is equal just to $\delta l$. Then one can easily show that the column equations are reduced to the following one:

$$FGQ^2 - FP^2 + G = 0 \tag{11}$$

where we denote $P = \partial T/\partial r, Q = \partial T/\partial l$. One has obviously

$$dT = Pdr + Qdl = -\left[G\left(Q^2 + F^{-1}\right)\right]^{1/2} dr + Qdl$$

that leads to equation

$$d\left\{T - Ql + \int \left[G(r)\left(Q^2 + F^{-1}(r)\right)\right]^{1/2} dr\right\} =$$
$$-\left\{l - \int \left[Q^2 G(r)/\left(Q^2 + F(r)^{-1}\right)\right]^{1/2} dr\right\} dQ \tag{12}$$

which is just the equivalent form of the column equation (11). A choice of a branch of square root is connected with a direction of motion of observers (inward or outward); our choice (with positive values of power 1/2 assumed) corresponds to motion *inwards*. Integrals in (12) are to be calculated with constant $Q$ of course.

There are two branches of (local) solutions of column equation (11), (12):
(**a**) *General branch*: $dQ \neq 0$. $Q$ can serve independent variable, at least locally. Then l.h.s. of (12) has to be equal to $\Omega(Q)'dQ$ for some (arbitrary) smooth function $\Omega(Q)$ and the implicit form of solution will be

$$T = -Q\Omega + \Omega' - \int \left[G/\left(F(Q^2 F + 1)\right)\right]^{1/2} dr, \tag{13.a}$$

$$l = -\Omega' + Q \int \left[GF/(Q^2 F + 1)\right]^{1/2} dr. \tag{13.b}$$

After integrating and eliminating of auxiliary variable $Q$ from eq. (13.a) by means of the resolving of (13.b) with respect to $Q$ one obtains the desirable column of the general branch.

(**b**) *Particular branch*: $Q \equiv K = $ constant. Here it follows from (12)

$$T = Kl - \int \left[G(r)F^{-1}(r)(K^2 F(r) + 1)\right]^{1/2} dr \qquad (14)$$

By the way, in this case eq. (11) falls into the ODE class itself and this formula could be inferred directly from (11) as well.

If the metric (9) is asymptotically flat in the sense that $F, G \to 1$ as $r \to \infty$ then

$$K^{-1} = v/[1 - v^2]^{1/2}.$$

Here $v = \lim(dr/dt) = $ constant is the "asymptotical speed" of column at infinity. The solution (10), (14) is *stationary*, i.e invariant with respect to transformation $t \Rightarrow t+$ constant (shift in time) accompanied by the transformation $l \Rightarrow l - $ constant$/K$ (shift of column), and this property distinguishes (14) in the whole set of solutions of (11). In the limiting case $K \to \infty$ ($v = 0$) one obtains *the static column*

$$l = \int \sqrt{G(r)} dr, \qquad t \text{ is arbitrary.}$$

2-dimensional static column positioned on a radial line can be obviously spread on the whole 4-dimensional space-time by means of the adding of equations $\varphi, \vartheta = $ constants. Clearly it coincides with the set of orbits of the static isometry.

Stationary columns turn out to be a generalization of inertial frames of reference (and reduce to them if $F = G = 1$ and space-time becomes flat) meanwhile the general branch of columns corresponds apparently to non-inertial motions of observers. Any arbitrarily moving observer can be included, at least locally, into some such column (and in this sense the capability criterion is verified). We shall see now how stationary (inertial) columns can be used for description of dynamics of point particles moving in the space-time under consideration.

3.2 DYNAMICS OF RADIAL MOTION IN CENTRAL SYMMETRIC STATIC SPACE-TIME

Let us consider a testing point particle moving in the space-time with metric (9) in accordance with equation

$$dt = -\mathcal{A}(r) dr \qquad (15)$$

for some known function $\mathcal{A}(r)$. Let its location coincide initially with one of observer from a stationary column (10), (14) distinguished by parameter $l$, and let after some short time coordinate $r$ of the particle be increased by a small increment $\delta r$. Then it finds itself in location of another observer distinguished by parameter $l + \delta l$ where

$$\delta l = \left\{\left[GF^{-1}\left(K^2 F + 1\right)\right]^{1/2} - \mathcal{A}\right\} \delta r / K.$$

Thus the particle has passed distance equal to $\delta l$. The time of its motion as measured by the first observer (with help of local synchronization principle, see Sec. **1**) can be found to equal

$$\delta\tau = \left\{\mathcal{A}\left(K^2 F + 1\right)^{1/2} - (GF^{-1})^{1/2}\right\}\delta r/K.$$

Then one can obtain the *speed* of particle with respect to column in hand

$$v \equiv \delta l/\delta\tau = \left\{\left[GF^{-1}\left(K^2 F + 1\right)\right]^{1/2} - \mathcal{A}\right\} / \left\{\mathcal{A}\left(K^2 F + 1\right)^{1/2} - (GF^{-1})^{1/2}\right\},$$

its *energy*

$$\begin{aligned}\mathcal{E} &\equiv \mu_0/(1-v^2)^{1/2} \\ &= \mu_0 \left\{\mathcal{A}\left(K^2 F + 1\right)^{1/2} - (GF^{-1})^{1/2}\right\} / \left\{K\left(F\mathcal{A}^2 - G\right)^{1/2}\right\}\end{aligned} \quad (16)$$

($\mu_0$ is the rest mass of particle), and *momentum*

$$\mathcal{P} \equiv v\mathcal{E} = \mu_0 \left\{\left[GF^{-1}\left(K^2 F + 1\right)\right]^{1/2} - \mathcal{A}\right\} / \left\{K\left(F\mathcal{A}^2 - G\right)^{1/2}\right\}. \quad (17)$$

Now one can calculate *a variation $\delta\mathcal{P}$ of momentum* on the infinitesimal path passed through by particle. Straightforward calculation yields

$$\delta\mathcal{P}/\delta\tau \equiv (\partial\mathcal{P}/\partial r)/(\delta\tau/\delta r) = \boldsymbol{f}_g + \boldsymbol{f}_s. \quad (18)$$

Here

$$\boldsymbol{f}_g = \mathcal{E} \cdot \tfrac{1}{2} K F' \left\{FG\left(K^2 F + 1\right)\right\}^{-1/2}, \qquad ('\equiv d/dr) \quad (19)$$

$$\boldsymbol{f}_e = -\mu_0 [FG]^{-1/2} \left\{F\mathcal{A}\left(F\mathcal{A}^2 - G\right)^{-1/2}\right\}'. \quad (20)$$

In according with relativistic version of Newton's law they are interpreted as the *forces* acting on particle [4]. Their classification turns out to be quite transparent. Indeed, the restriction to $\boldsymbol{f}_e$ to vanish is precisely the geodesic equation for motion in accordance with eq. (15) in metric (9) and, hence, in case of *free* particle motion when the only gravitational force may exist it shall coincide with $\boldsymbol{f}_g$. This interpretation of $\boldsymbol{f}_g$ clearly retains for any other motion of particle as well.

---

[4]) The expression for the force acting on a test particle in stationary gravitation field derived in physical terms has been suggested in [9], p. 252-253, eq. (3). No general concept similar to column was introduced there but the congruence of orbits of the symmetry transformation group is in fact a partial example of a column. On the contrary, the approach being developed by the authors of [10] (see also [11] and references therein) has no relation to our comprehension of the force notion.

Eq. (19) reveals all the features desirable for gravitational force expression. It is proportional to the *total* mass of particle (not merely to a rest one!) and this is the only entry of properties of particle or characteristics of its motion into the force expression (19). The latter fact confirms a validity of *equivalence principle* for the case in hand. Furthermore, one may suggest the following expression for *the strength of gravitational field* measured with respect to chosen column:

$$\mathcal{F} = \tfrac{1}{2} K F' \left\{ F G \left( K^2 F + 1 \right) \right\}^{-1/2}, \qquad (21)$$

the corresponding "charge" being the total mass (equal to energy $\mathcal{E}$ since we assume $\mathbf{c} = 1$). In the most physically interesting Schwarzschild case we have a generalization of Newton's law

$$\mathcal{F}_{\text{Sch}} = \frac{m}{r^2} \left[ \frac{1}{1-v^2} - \frac{2m}{r} \right]^{-1/2}. \qquad (22)$$

Here $m$ is the Schwarzschild mass (and $v$ is the asymptotic speed of column, see Sec. **3**.1 above) [5]). In particular one has for *the static column*

$$\mathcal{F}_{\text{Sch}}^{st} = \frac{m}{r^2} / \sqrt{1 - 2m/r}.$$

It is worthwhile to emphasize that introduction of consistent notions of gravitational force and strength of gravitational field would be extremely important because, as to the author's opinion, they are the indispensable tools for attack to the problem of gravitational field energy. Still, we give here only the simplest example of such sort calculation.

An expectable but nevertheless pleasant fact is that the case of zero rest mass particles (photons) admits almost the same consideration. Here only geodesic world lines make sense and the second term in (18) vanishes automatically. Although the energy cannot now be expressed by means of the first equality in (16) we may assume it to be proportional to the frequency associated with photon which in turn can be realized as a frequency of homogeneous series of several photons (null geodesics) registered by observer. Then the same way of calculations leads precisely to the same expression (19). Thus photons are undergone the same gravitational force as non-zero-rest-mass particles are (which is another manifestation of the principle of equivalence). And it is not surprising that as a result of the action of gravitational force they *deflect* from the possible "straightest" way (geodesic of the physical space).

As to the formula (20), it determines just the *external* ( non-gravitational) *force* acting on particle. The most simple and important relevant example is provided by motion of a charged particle in the Reissner-Nordström space-time. Here

$$F = G^{-1} = 1 - 2m/r + q^2/r^2,$$

---

[5]) It is worthwhile to note that for $v>o$ the stationary column will *penetrate* through the horizon of Schwarzschild black hole and the forces reveal no peculiarities on it (but diverge somewhat later). Besides, the distance from horizon to observer moving *inward* is finite.

$$\mathcal{A} = JF^{-1}[J^2 - \beta^2 F]^{-1/2}, \quad J = 1 - \chi/r, \quad \chi = \beta eq/\mu_0$$

and $m, q$ and $\mu_0, e$ are the masses and charges of central body and particle respectively, $\beta$ is a constant of motion. Evaluating expression (20), just the classical Coulomb's formula results:

$$\boldsymbol{f}_e = -\frac{eq}{r^2}. \tag{23}$$

Thus we have obtained in particular the physical (measurable) strength $\mathcal{F}_e = q/r^2$ of electric field of a charged black hole. (The problem of the "physical interpretation" of the strengths of non-gravitational fields is of course of the same importance as the case of gravitational field but it is not generally speaking the subject of this work.)

It is worthwhile to notice for clearness that all the scalar functions (17)-(23) are in fact the (only) components of the corresponding vectors with respect to orthonormal frame $\{\frac{\partial}{\partial l}\}$ belonging to the (Euclidean 1-dimensional) physical space.

These simple examples seem already to be sufficient to prove to some extent a physical reasonability of the approach under consideration.

### 3.3 FORCE OF INERTIA IN FLAT SPACE-TIME

The column equations have nontrivial solutions even in case of flat space-time. For the sake of simplicity we consider here a special case which corresponds to effectively 1-dimensional motion of observers. Let their set can be foliated into a family of parallel planes in such a way that every observer belongs to a certain plane and each plane may move in direction normal to it ($z$-direction) only. Let some chosen observer move in according with equation

$$z = \zeta(t). \tag{24}$$

where the restriction $|\zeta'| < 1$ is to be observed. The column "leaning" on this observer under conditions assumed is unique and described by equations

$$\zeta'(\sigma)l = \left[1 - (\zeta'(\sigma))^2\right]^{1/2}(t - \sigma), \tag{25a}$$

$$z = \zeta(\sigma) + l\left[\frac{(t-\sigma)^2}{l^2} + 1\right]^{1/2}, \tag{25b}$$

$$x = l_x, \quad y = l_y.$$

Here $\sigma$ is auxiliary variable which is to be expressed as a function of $t$ and $l$ from eq. (25a); it is possible to do this, at least locally, if

$$\mathcal{G} \equiv \left[1 - (\zeta')^2\right]^{3/2} + l\zeta'' \neq 0.$$

It is clear that $\mathcal{G}$ can vanish only "occasionally" in separate points and, moreover, only for sufficiently large $l$. Variables $l_x, l_y, l_z \equiv l$ enumerate observers, the physical metric being Cartesian with respect to them. In particular, they measure distances along corresponding axes.

Eqs. (25) imply a general solution $z = Z(t, l)$, $x = l_x$, $y = l_y$ of column equation under conditions assumed and one can prove that $\lim Z(t, l) = \zeta(t)$ as $l \to 0$, provided, for example, $|\zeta'| \leq 1 - \delta, \delta > 0$. The case of 'uniformly accelerated' column is realized as an evident particular case of the above formulae.

An analysis of motion of free particles with respect to column (24) yields the following expression for strength of force acting on particle:

$$\mathcal{F}_i = -\zeta''/\mathcal{G}, \qquad (26)$$

corresponding "charge" being the total mass. It is really the vector of $l$-direction of course. Due to our *a priori* knowledge of absence of the true gravitational force (space-time is flat) we can assert that (26) is just *the strength of force of inertia.*

Next let us consider a resolvability of eq. (24a) with respect to $\sigma$ in more details. To that end, we recast it to the form

$$(t - \sigma)/l = \Psi(\sigma) \qquad \text{where } \Psi = \zeta' \left[1 - (\zeta')^2\right]^{-1/2}$$

and interpret its solution as intersection of graphs of the linear function of $\sigma$ $L(\sigma) = (t-\sigma)/l$ and the function $\Psi(\sigma)$. It is evident that for small $l$ intersection always exists and is unique. But *if the motion* (24) *is non-inertial*, i.e. if $\zeta' \not\equiv$ constant $\Leftrightarrow \Psi \not\equiv$ constant, then increasing $l$ one will inevitably achieve the point where $L$-line will become tangent to $\Psi$-graph. In vicinity of such a point the solution of eq. (25a) takes the form $\sigma = \sigma_0 + \delta\sigma + o(\delta\sigma)$ where

$$\delta\sigma = \pm \left[\frac{2(t_0 - t)}{l_0 \Psi''(\sigma_0)}\right]^{1/2}.$$

Thus, formally, a value of coordinate $t$ cannot surpass $t_0$. This means in fact that the function $\sigma(t, l)$ is singular in vicinity of arguments $t_0, l_0$ that corresponds to a failure in that point of the smoothness of column (a 'catastrophe').

It is important that some singularities are peculiar to *any* column of considered type *excluding inertial ones*. We have presented above the proof (although not quite rigorous) of this fact for the case under consideration. Examples proves that the larger acceleration of column observers, the less space-time region covered by column. Thus it seems reasonable to put forward the following

**Conjecture.** *Let the set of all columns be partially ordered by relation of inclusion of maximal space-time areas that are covered by columns and where they are free of singularities. Then maximal elements of this set correspond to inertial columns.*

One may conjecture that this principle of separation of inertial columns suits also for space-times without asymptotically flat regions where one has no other guides enabling to distinguish inertial motion.

### 3.5 TWO-DIMENSIONAL COLUMN IN FLAT SPACE-TIME

Next level of complexity is 3-dimensional column corresponding to a 2-dimensional motion of family of observers. Important feature peculiar to this case is a permissibility of *rotation*. A surprising fact is that the corresponding column equations in *flat* space-time admit a complete solution in quadratures involving arbitrary functions and thus can be analyzed in all details. Unfortunately, it seems impossible to describe this general solution here and we give only a partial example.

Let us express the standard coordinates $t, x, y$ in 3-dimensional Minkowski space-time in terms of parameters $p, \chi, \varphi$ as follows:

$$t = p \sinh \chi, \quad x = p \cosh \chi \cos \varphi, \quad y = p \cosh \chi \sin \varphi.$$

Let us assume further that $p$ is constant along column world lines, *i.e.* it could serve as a parameter enumerating observers. We denote the second enumerating parameter $q$. It turns out that one may identify $q$ and $q + 2\pi$, so $q$ is an angle variable; furthermore $q$ may be considered as a function of $\chi$ and $\varphi$: $q = q(\chi, \varphi)$. Then one can prove that the column equations are satisfied if

$$\left(\frac{\partial q}{\partial \varphi}\right)^2 - \left(\frac{\partial q}{\partial \chi}\right)^2 \cosh^2 \chi = \cosh^2 \chi. \tag{27}$$

General solution of eq. (27) consists of two branches. The first of them corresponds to restriction $\partial q / \partial \varphi \equiv$ constant and can be obtained by a simple quadrature. The second branch of solutions is less trivial. It is described in implicit form by formulae

$$\cos(q + \Phi - \nu \Phi') = \cosh \chi \cos(\varphi - \Phi') = \left[\frac{\nu^2 - \cosh^2 \chi}{\nu^2 - 1}\right]^{1/2},$$
$$\nu \sin(q + \Phi - \nu \Phi') = \cosh \chi \sin(\varphi - \Phi'). \tag{28}$$

Here $\nu$ is auxiliary variable subjected to elimination and $\Phi = \Phi(\nu)$ is arbitrary function. The physical metric is flat

$$dl^2 = dp^2 + p^2 \, dq^2,$$

$p$ playing role of a radius with respect to it (and this is a *rigorous* and *unique* its interpretation).

It should be noted that there are 3-dimensional columns in flat space-time with non-flat (although rather specific) physical metric as well.

Possible physical interpretations of solution (28) remain open and hence it represents rather a mathematical result for present.

### 3.6 UNIFORM ROTATION IN SPHERICALLY SYMMETRIC STATIC SPACE-TIME

Our last example of 2-dimensional column is perhaps less mathematically diverting but certainly more physically meaningful. Let us consider 3-dimensional space-time endowed with metric

$$\boldsymbol{ds}^2 = F(r)dt^2 - G(r)dr^2 - H(r)d\varphi^2. \qquad (29)$$

It coincides with restriction of a general static spherical symmetric metric to equatorial plane $z = 0$. Uniformly rotated column is determined by equations

$$\varphi = \omega t + q, \quad r = P(p). \qquad (30)$$

Here $\omega$ is some constant, function $P$ is defined by means of equation $p = P^{-1}(r) = \int \sqrt{G(r)} dr$, $p$ and $q$ are the parameters identifying observers (the latter has to be identified modulo $2\pi$). The physical metric can be easily shown to be

$$\boldsymbol{dl}^2 = dp^2 + \frac{H(r)F(r)}{F(r) - \omega^2 H(r)} dq^2. \qquad (31)$$

Calculation yields the following vector of combined gravitational-inertial force:

$$\boldsymbol{f} = \frac{1}{\sqrt{G}(F - \omega^2 H)} \left\{ \frac{1}{2} \left( F' - \omega^2 G' \right) \mathcal{E} \frac{\partial}{\partial p} + \omega (FH)^{1/2} \left( F'/F - H'/H \right) \star \mathcal{P} \right\}. \qquad (32)$$

Here $\mathcal{P}$ is the vector of particle's momentum, $\star$ denotes dualizing operator (anti-clockwise rotation by $\pi/2$ in our case). One may consider the vector $\frac{\partial}{\partial p}$ to be of a radial direction if desirable but the real sense of such a statement should be verified by the eq. (31) and may depend on the functions $F, H$ and $P$ (*i.e.* essentially $G$).

We have no undoubted principles enabling us to separate gravitational and inertial constituents of the force (32) and even an existence of such a principle is problematical (is usually refused, to be more exact) although in some cases one may definitely assert that the manifested force is mainly of inertial (or, inversely, gravitational) nature. Nevertheless, the following suggestion seems to be rather likely:

We assume

$$\mathcal{F}_g = -\frac{1}{2} \frac{F'}{FG^{1/2}} \cdot \frac{\partial}{\partial p} \qquad (33a)$$

to be *the strength of gravitational field* (cf. eq. (21)),

$$\mathcal{F}_{cf} = \omega^2 \frac{H}{2G^{1/2}} \cdot \frac{H'/H - F'/F}{F - \omega^2 H} \cdot \frac{\partial}{\partial p} \qquad (33b)$$

to be *the strength of centrifugal force* and

$$\mathcal{F}_{\text{Co}} = -2\omega \frac{F^{1/2} H^{1/2}}{2G^{1/2}} \cdot \frac{H'/H - F'/F}{F - \omega^2 H} \cdot \star \qquad (33c)$$

to be *the operator of Coriolis' strength*. Then the corresponding resulting force will coincide with (32). (Of course, only the later expression is identified unambiguously, at the same time only the sum of previous two ones can be strictly speaking distinguished.)

The magnitude $|\mathcal{F}_g|$ of strength (33.$a$) does not depend on the frequency parameter $\omega$ which in particular characterizes the motion of observers in direction normal to the direction of strength and this properties seems to be quite reasonable in case of central symmetric gravitational field.

Under separation (33) the gravitational strength is everywhere nonzero but inertial forces vanish on the null circular orbit $r = 3m$ of Schwarzschild space-time and reverse their 'direction' when one passes it [6]. The gravitational strength diverges on the black hole horizon.

## 4 DISCUSSION

The examples given in preceding section demonstrate a fruitfulness and plausibility of approach based on the notion of column. It seems difficult to object against that measurements of space and time relationships with respect to column have direct physical meaning, *provided the column exists*.

One may argue however that cases of existence of column could be rather rare and, furthermore, that they are likely to be connected with symmetries of space-time (at least, all the examples demonstrated in Section **3** referred to highly symmetrical geometries); thus the approach suggested can not be regarded to be very profound.

Indeed, a *strong* column may not exist in a generic situation but the *principle of reference* mentioned in Section 1 remains still valid. We reproduce it here in the following form: *All the space and time relationships are to be refereed to a family of observers moving in such a manner that their "media" should be locally as stiff as possible, at the same time a total control over inevitable intrinsic motions of its constituents is to be observed.*

The weak column notion (see Sec. **1**) is a possible realization of this principle in generic situation. I have obtained an example of non-trivial 3-dimensional weak column which will be published elsewhere. It corresponds to a spread of radial column described in subsection **3**.1 onto the whole equatorial plane of central symmetric static space-time. Additionally, this facility enables one to spread the asymptotic Lorentz group from almost flat infinity to curved inner region. Calculation of dynamical field quantities (gravitational and inertial strengths) yields quite reasonable results as well although it turns out to be insufficient to remain in frames of Riemannian geometry when analyzing relationships *in physical space*.

In this example, one does not take advantage of of space-time symmetries since they are effectively broken by the motion of column's observers not following the symmetry. As a result the picture of fields observed by column is non-stationary and almost non-symmetric and this is by no means a coordinate or gauge-like effect. There are no doubts that the last simplifying restriction – diminished dimension – can be removed and a complete 4-dimensional proper weak column can be constructed. Thus the approach in consideration does not exploit merely the "resides" of a flat space-time (symmetries) in specifically curved ones; it is applicable to not only symmetric geometries.

---

[6]) The latter property of the forces of inertia has been suggested (in completely different context, however) in the work [11] and cited therein ones.

On the other hand, if the "strong" columns were obliged for their existence to the "reside" of symmetries inherited by space-time from Minkowski space, the more direct significance should be attached to the physical consequences that are implied by them although in such a case they represent only particular cases of gravitational fields.

Finally, we may assert that the approach introduced in the work has a clear physical meaning and certainly is worthy of further analysis and development.

**Acknowledgment**. I am grateful to the *Graduertenkolleg* "Scientific computing" (Köln – St. Augustin, Nordrhein-Westfalen) for financial support and to the Institute of Theoretical Physics of the University of Cologne for hospitality.